# Reconstructing the late accretion history of the Moon

Meng-Hua Zhu[1]∗, Natalia Artemieva[2], Alessandro Morbidelli[3], Qing-Zhu Yin[4], Harry Becker[5], Kai Wünnemann[5,6]

[1]State Key Laboratory of Lunar and Planetary Sciences, Macau University of Science and Technology, Taipa, Macau;

[2]Planetary Science Institute, Tucson, USA;

[3]Département Lagrange, University of Nice–Sophia Antipolis, CNRS, Observatoire de la Côte d'Azur, Nice, France;

[4]Department of Earth and Planetary Sciences, University of California at Davis, Davis, CA 95616, USA;

[5]Institute für Geologische Wissenschaften, Freie Universität Berlin, Berlin, Germany;

[6]Museum für Naturkunde, Leibniz Institute for Evolution and Biodiversity Science, Berlin, Germany

∗correspondence to: mhzhu@must.edu.mo

**The importance of highly siderophile elements (HSEs) to track planetary late accretion has long been recognized. However, the precise nature of the Moon's accretional history remains enigmatic. There exists a significant mismatch of HSE budgets between the Earth and Moon, with the Earth disproportionally accreted far more HSEs than the Moon did[1]. Several scenarios have been proposed to explain this conundrum, including the delivery of HSEs to Earth by a few big impactors[1], the accretion of pebble-sized objects on dynamically cold orbits that enhanced the Earth's gravitational focusing factor[2], and the "sawtooth model" with much reduced impact flux before ~ 4.10 Gyr[3]. However, most of these models assume a high impactor retention ratio $f$ (fraction of impactor mass retained on the target) for the Moon. Here, we performed a series of impact simulations to quantify the $f$-value, followed by a Monte Carlo procedure enacting a monotonically decaying impact flux[4], to compute the mass accreted into lunar crust and mantle over their histories. We found that the average $f$-value for the Moon's entire impact history is about 3 times lower than previously estimated[1-3]. Our results indicate that, to match the HSE budget of lunar crust and mantle[5,6], the retention of HSEs should have started 4.35 Gyr ago, when most of lunar magma ocean was solidified[7,8]. Mass accreted prior to 4.35 Gyr must have lost its HSE to the lunar core, presumably during the lunar mantle crystallization[9]. The combination of a low impactor retention ratio and a late retention of HSEs in the lunar mantle provide a realistic explanation for the apparent deficit of Moon's late accreted mass relative to the Earth.**

(Main text)

The Moon was formed by accreting debris of mantle material from the target proto-Earth and impactor[10,11], initially deficient in HSEs owing to their strong affinities to metal relative to silicates. The ensuing differentiation and core formation depleted HSE from the early silicate Moon[12,13]. Subsequently, the Moon has undergone a long-term bombardment after its formation[3,4], and a significant amount of exotic materials was delivered to the Moon. The accretion of extra-lunar materials after the cessation of Moon's core formation can be traced by the enrichment of HSEs in the lunar crust and mantle. The late-arriving impactors, with relative HSE abundances similar to chondritic meteorites[13], replenished the lunar mantle HSE content before the formation of lunar crust[5,14]. After solidification of a thick lunar crust, only larger impactors could penetrate into the mantle. As a consequence, in most cases the impactor materials are mixed into "pristine" lunar crust, which was initially characterized by extremely low HSE abundances[6,15]. Estimates from the HSE contents in mantle-derived magmatic rocks suggests ~ $1.7 \times 10^{19}$ kg of material with a chondritic bulk composition accreted in the lunar mantle[5,14], whereas ~ $0.45\text{-}1.0 \times 10^{19}$ kg of chondritic material were mixed into the crust, as estimated from the HSEs contents in ferroan anorthosites[6] and bulk regolith samples[16] (see Methods for details). These accreted masses provide significant constraints on the Moon's impact flux[3,9]. However, the timeline and detailed accretionary process are poorly constrained. Intriguingly, the total mass accreted to the Moon (~$2.1\text{-}2.7 \times 10^{19}$ kg) is about three orders of magnitude smaller than that accreted to the Earth, as inferred from HSEs in Earth's mantle (~ $2.0 \times 10^{22}$ kg, *ref*. 14). This contrasts with the ratio of ~ 20 in collisional cross-sections, after gravitational focusing is taken into account[1-3]. Several scenarios have been proposed to explain this significant imbalance[1-3,17], however, its origin is still enigmatic.

Interpretation of the late accreted masses as constraints on the accretionary process requires a quantitative evaluation of the fraction of the impactor material accreted upon impact (hereafter called *retention rat*io, *f*). Impactors hitting the Earth are expected to be almost fully accreted due to the Earth's high gravity (see Methods, *refs*. 1, 3, 17). However, how the retention ratio varied for individual impacts on the Moon is unclear[18]. A rough assumption of *f* ~ 0.5-0.6 has been frequently used[1,3]. In order to quantitatively investigate the impactor retention ratio on the Moon, we performed a suite of oblique impact simulations (impactor diameter $d$ = 10-560 km, impact velocity $v$ = 10, 15, and 20 km s$^{-1}$, and impact angle $a$ = 20°-80° with respect to the lunar surface) on a spherical Moon with the iSALE-3D shock physics code[19]. We recorded the fractional mass of the impactor that is not ejected or is ejected with a velocity below the Moon's escape velocity (~ 2.4 km s$^{-1}$) and considered these impactor materials to be retained by the Moon (see Methods). Our simulations show a significant variation of impactor retention ratios on the Moon (see Fig. 1). The retention ratio decreases exponentially with increasing impactor-to-target size ratio ($x$) for impacts with fixed velocity and angle: high-angle impact delivers a larger fraction of impactor material than a low-angle impact; large impactors deliver proportionally less fraction of material than small ones (see Methods).

With these individual impactor retention ratios, we conducted a Monte Carlo simulation assuming an impact flux that follows the crater production function advocated in (4), the prescribed impact velocities of (20) and isotropic impact angles[21] to reproduce the Moon's impact history and assess the mass accreted into the lunar crust and mantle, respectively. In detail, the impactor size-frequency distribution was converted from the crater production function of the Moon[4] with crater-to-projectile-size scaling laws[22] and was then used to generate random impactors forming craters up to the observed basin-sized structures (see Methods). Whether the *retained* impactor material adds to the lunar crust or the mantle depends on the penetration depth of the impactor relative to the crustal thickness and the fraction of the impactor material that is expelled from the growing crater and deposited on the surface. If the transient crater depth is larger than the average crust thickness of the Moon, the fraction of the *retained* impactor emplaced within the transient cavity is considered to deliver its material (and HSEs) to the mantle, while the fraction deposited beyond the transient crater is considered to mix with the crust. Otherwise, the *retained* material is considered to mix with the crust (see Methods).

For simplicity, we assume a constant crustal thickness (e.g., 34 - 43 km, *ref.* 23) from the time of crust formation[24] until present day. We considered the time at which the mantle and crust started to retain HSEs as a free parameter, and determined the value that allows the simulation to match the observed HSE budgets in the lunar mantle and crust, respectively.

We repeated the Monte Carlo procedure millions of times to address the stochastic variability intrinsic in the bombardment process. A key result is that the Moon's average impactor retention ratio depends on the time-interval considered (see Fig. 2). Early on, more large-size impactors lead to a lower average retention ratio because large impacts deliver a smaller fraction of impactor material to the Moon (see Fig. 1). The average retention ratio integrated over the Moon's impact history since 4.46 Gyr, the presumed earliest formation time of lunar crust[24] is ~ 0.20 (see Fig. 2), which is about 3 times lower than the range of 0.5-0.6 that was frequently assumed previously[1,3,9].

For the accreted mass from our simulations to be consistent with that inferred from the HSE concentrations in the lunar mantle, the required time at which the lunar mantle should have begun to retain impactor HSEs is 4.35 Gyr, namely ~ 150 Ma after the presumed lunar formation event[25]. In the model, the mantle accretes ~ $1.70 \times 10^{19}$ kg of impactor material since 4.35 Gyr, when the lunar magma ocean (LMO) was mostly solidified and a thick crust was fully formed[26,27] (see Fig. 3); ~ 85% of this mass is typically accreted between 4.35 Gyr and 4.15 Gyr. The mantle accretes significantly more material, ~ $3.0 \times 10^{19}$ kg, before 4.35 Gyr. This suggests that the early mantle of the Moon could have had substantially higher abundances of HSEs compared to the current Moon. Impactor material accreted to the mantle before 4.35 Gyr did not leave behind any record in current HSE budget either because it was trapped in the deep mantle[5] or because the HSEs partitioned into the core[28] during the mantle crystallization and overturn[7, 9]. This result supersedes previous assumptions that the HSEs in the lunar mantle accumulated early, prior to the formation of a thick crust (e.g., *refs*. 5, 6, 14). Contrary to previous assumptions that the late-occurred impacts did not contribute to the HSE budget of the lunar mantle due to the existence of a thick crust, we find that large impacts can penetrate the thick crust and deliver impactor material into the mantle, even when the crust has been well developed and cooled. However, the majority of these replenishments by large impacts happened before ~ 3.85 Gyr. After this date the delivery of impactor material to the mantle is small (see Fig. 3). This is because there are only few impacts that can penetrate through the thick lunar crust after 3.85 Gyr[29].

If the lunar crust retained impactor material since 4.46 Gyr - its estimated earliest formation time[24], the total accreted mass in the lunar crust could be up to ~ $1.0 \times 10^{19}$ kg (see Fig. 3). For any later retention time, but before 4.35 Gyr, the total accreted mass in the crust is within the range of $0.45\text{-}1.0 \times 10^{19}$ kg inferred from the estimated crustal HSEs budget (e.g., *refs*. 6, 16 and Methods). Therefore, this mass range ($0.45\text{-}1.0 \times 10^{19}$ kg) may represent the lower and upper limit of the material accreted to the lunar crust, respectively. It indicates that the lunar crustal HSE budget may represent the accretion history since the crust's earliest formation time, at least since 4.35 Gyr when the LMO was mostly solidified[7, 26, 27].

The estimated total mass of impactors hitting the Moon since 4.50 Gyr is ~ $3.7 \times 10^{20}$ kg (~ $6 \times 10^{-5}$ $M_{\oplus}$) (see Fig. 3). This mass is an order of magnitude higher than previous estimates based on the lunar HSE budget ($5 \times 10^{-6}$ $M_{\oplus}$, e.g., *ref.* 3), but in agreement to the recent estimation from dynamical considerations[9]. This suggests that the Moon underwent a much more intense early bombardment than previously considered[3]. Our simulations indicate that the impactors should have produced ~ 300 basins ($D > 300$ km) during the lunar impact history: ~ 200 basins before 4.35 Gyr, ~ 90 basins between 4.35

Gyr and 4.15 Gyr, and ~20 basins from 4.15 Gyr to the present day. This number is ~ 3-8 times higher than the number of basins (~40-90) estimated from certain and uncertain structures observed on the lunar surface[30, 31]. However, impacts that occurred before 4.35 Gyr, during the main phase of LMO crystallization[7], should have failed to produce long-lasting structures because of the low viscosity of the warm crust and mantle. In addition, owing to the viscous relaxation of the target, basins formed just after LMO solidification (~ 4.35 Gyr) can exist for ~ 100 Ma at most[32]. Consequently, about 40-50 of the 90 basins that formed between 4.35 Gyr and 4.15 Gyr (~ 200 Ma) were probably erased, whereas the others partially relaxed to some extent. Later formed basins were preserved and remain detectable to the present day. Therefore, according to our model and given basin retention and degradation processes[32], only 50-70 basins in total should be visible, in agreement with the number of basins (~ 40-90) observed or inferred on the lunar surface[30, 31]. These basins are the only remnants of the heavy bombardment history of the Moon.

With previous assumptions of an average impactor retention ratio of 0.5-0.6 (Fig. 2, and *refs*. 1, 3), the *Neukum et al.* impact flux[4] predicted a mass of late accreted material an order of magnitude higher than the mass inferred from the HSEs in the lunar mantle and crust[3]. With the reduced value of the impactor retention ratio $f$, the amount of accreted material still exceeds that inferred from the HSE budget, although only by a factor of ~ 3. A sawtooth-like impact flux[3], with a surge in the projectile population at the time of a presumed lunar cataclysm (Extended Data Figure 4), was proposed to reconcile the impact history of the Moon and the amount of accreted material during the last 4.50 Gyr inferred from the HSEs. However, if the lunar mantle retained HSEs only since 4.35 Gyr, the conceptual sawtooth-like cataclysm scenario[3] is not needed anymore to explain the Moon's impact flux (see Methods).

The good agreement between the time at which the lunar mantle started to retain the accreted HSEs (~ 4.35 Gyr) and the age records from lunar crustal rocks (e.g., *ref.* [26]) indicate that the HSE budget of the lunar mantle possibly depends on the time of crystallization of the LMO. Our results support an extended solidification time scale of the LMO of ~150-200 Ma after lunar formation[7]. Our results suggest that HSEs accreted during or before crystallization of the LMO were mostly transported to the lunar core during late-stage metal and sulfide segregation[9].

The estimated total mass of impactors hitting the Moon since 4.50 Gyr (~ $3.7 \times 10^{20}$ kg) is about a factor of 50 times lower than the mass accreted to the Earth (~ $2.0 \times 10^{22}$ kg, *ref.* [14]), assuming that the impactors hitting the Earth were fully accreted [1, 3, 9, 17] and retained in the terrestrial mantle owing to the fast crystallization of its magma ocean[33]. The ratio of collisional cross-sections is ~ 20 implying that the alleged large imbalance of late accreted mass between the Earth and Moon[1] no longer exists. The disproportional mass accreted on the Earth and Moon inferred from the HSEs[1,13] is most probably a consequence of different impactor retention ratios coupled with different time scales of magma ocean solidification on these bodies.

**Acknowledgements**

We thank J. M. D. Day and R. J. Walker for useful discussions. We acknowledge the developers of iSALE ([www.isale-code.de](www.isale-code.de)), in particular to Dirk Elbeshausen for developing the iSALE-3D. M.Z. is supported by the Science and Technology Development Fund of Macau (079/2018/A2). K.W., H. B., N.A., and M. Z. are funded by the DFG grant SFB-TRR 170 (A4, C2), TRR-170 Pub. No. 55. Q.-Z.Y. is funded by NASA Emerging Worlds Program (NNX16AD34G).

**Author Contributions**

M. Z. conceived the idea and performed the impact simulations. N. A. performed the Monte Carlo modeling. M. Z., A. M., Q.–Z.Y, H. B., and K. W. interpreted the results. All authors contributed to the discussion of the results and wrote the manuscript.

**Additional Information**

Reprints and permissions information is available at www.nature.com/reprints. The authors declare no competing financial interests. Readers are welcome to comment on the online version of the paper. Correspondence and requests for materials should be addressed to M.Z. (mhzhu@must.edu.mo).

**Figure 1 | Impactor retention ratios (*f*) as a function of the impactor-to-target size ratio, impact angle and velocity. a**, Impactor retention ratio for the case of impact velocities of $v$ =10 km s$^{-1}$. **b** and **c**, as in a, but for $v$ = 15 and 20 km s$^{-1}$. The impactor-to-target size ratio refers to the impactor diameter relative to that of the Moon. In our simulations, the assigned impact angles (*a*) were 20º, 30º, 45º, 60º, 70º, and 80º relative to the horizontal surface. The points represent the values derived from the iSALE modeling, whereas each line shows a fit with an exponential function of $f = a*\exp(-b*x)$. Here, $x$ is the ratio between impactor diameter and Moon's diameter. The parameters (*a*, *b*) of the fitted exponent function are shown in Extended Data Table 2.

**Figure 2 | Average impactor retention ratio as a function of the time.** The Moon's average impactor retention ratio *f* (solid green squares) is defined as the ratio between the total mass accreted to the Moon and total impactor mass hitting the Moon. For example, data point at 4.35 Gyr represents the average impactor retention ratio for impacts occurring cumulatively from 4.35 Gyr to the present-day. In this work, the average impactor retention ratios are within 0.20-0.35 (red-hatched area) for a start time between 4.46 Gyr and 3.50 Gyr, similar to the range predicted (0.16-0.32 for the entire impact history of the Moon) in *Schlichting et al.*[2], but is ~ 2-3 times lower than the value considered in *Bottke et al.*[1] and *Morbidelli et al.*[3] (*f* = 0.50 or 0.60, the black-hatched area, derived from the oblique simulation of small impactors at an angle of 45-degree[18]). The average impactor retention ratios for the lunar crust, defined as the ratio between the total mass accreted to the crust with a thickness of 34 km (orange circles) or 43 km (gray circles) and total impactor mass hitting the Moon, are also computed.

**Figure 3 | Cumulative impactor mass distribution as a function of time.** Shown in blue are cumulative impactor mass hitting the Moon, in purple are the cumulative mass being accreted on the Moon since different starting times (between 4.5 Gyr to 3.5 Gyr

ago) to the present-day. The assumed crustal thickness is 34 km (**a**) and 43 km (**b**), respectively. The cumulative mass accreted to the crust (orange) and the mantle (green) of the Moon are estimated separately. Each data point represents the average of millions of Monte Carlo simulations. We assume a projectile density of 3,000 kg m$^{-3}$. The horizontal black and green lines represent the mass accreted to the silicate part of the Moon (2.20 x 10$^{19}$ kg) and mantle (1.70 x 10$^{19}$ kg) inferred from HSE budgets[5], respectively, whereas the horizontal brown lines mark the lower (0.45 x 10$^{19}$ kg) and upper (1.0 x 10$^{19}$ kg) limits for the mass accreted to the lunar crust, inferred from HSE as well[6,16]. The vertical black line represents the time of 4.35 Gyr when the masses added to the crust and mantle from the Monte Carlo simulation reach the values (horizontal lines) estimated from the HSEs. The lower (34 km) and upper (43 km) limits for the average crustal thickness used in this work are from the observations of GRAIL mission[23]. As lunar crust may have formed (~ 4.46 Gyr, *ref. 24*) soon after the Moon's formation[25], for simplicity, we assume in our calculations that the global crustal thickness remained constant during the lunar history. Assuming an evolving crustal thickness from a few kilometers soon after Moon formation to the present thickness at ~ 4.46 Gyr would not affect the results.

## METHODS
**Impactor mass accreted to the lunar crust and mantle derived from the HSE.**

The masses of impactor martial accreted to the crust and mantle of the Moon are estimated from the concentrations of HSEs in the lunar samples (for example, *refs. 5, 6, 34, 35*). The HSE concentrations in the lunar mantle is robustly estimated, regardless of sulfide or metal saturation in the lunar mantle[36]. According to the Os concentration (~100 pgg$^{-1}$) of mantle-derived melts (for example, mare basalts and pyroclastic glasses, *refs. 5, 34, 37*) in lunar samples, it is conservatively estimated that ~ $1.70 \times 10^{19}$ kg material with a chondritic bulk composition was added to the lunar mantle[34], assuming that the mass of lunar mantle is ~ $6.9 \times 10^{22}$ kg[38] and that late-accretion-delivered bodies possess an average Os concentration of 660 ng g$^{-1}$, similar to chondrites[14].

For the mass of late accreted material stored in the crust, estimates based on HSE contents have varied considerably. *Day et al.*[6] analyzed the concentrations of HSEs for the pristine crustal rocks (60025, 62255, 65313) of the Moon and found that the floatation-derived anorthositic crust has a typical Os concentration of 1.4 pg g$^{-1}$. In contrast, impact melt breccias and bulk regolith samples contaminated with impactor material have Os concentrations mostly within 5-15 ng g$^{-1}$ (see *refs. 39-43*). The extent of contamination of the lunar crust by impactors is uncertain, in particular at greater depth. Estimates range from ~ 10% of crustal mass (corresponding to the uppermost ~ 5-10 km of crust, for example, *ref. 6*) to a roughly uniform mixture of impactor material with the crust (e.g., *ref. 2*). With these assumptions, a mass of $0.4 \times 10^{19}$ kg[16], but no more than $14 \times 10^{19}$ kg[2] of material is thought to have been accreted to the lunar crust. However, the high-end estimate by *Schlichting et al.*[2] implies uniform Os and Ir distribution throughout the lunar crust, which is unrealistic.

The observations of Gravity Recovery and Interior Laboratory (GRAIL) mission suggest an average crustal thickness between 34 km and 43 km[23]. With a bulk density of 2,500 kg m$^{-3}$ (*ref. 23*), the mass of the lunar crust is ~ 3.2 - 4.1 x 10$^{21}$ kg. Assuming the crust has an Os concentration within 5-15 ng g$^{-1}$ and ~ 10% of lunar crust was contaminated by the impactor material (e.g., *refs. 6, 37, 43*), we recalculated the mass of chondritic material accreted to the lunar crust. Our calculation indicates that chondritic material with a mass between ~ 0.45 x 10$^{19}$ kg and ~ 1.0 x 10$^{19}$ kg was added into the lunar crust, consistent with the early estimates (e.g., *refs. 6, 16, 35*). In this work, we use these values as lower and upper limits to estimate the time since which significant masses of impactor material were retained in the lunar crust.

**Impactor retention ratio on the Moon.**

In order to investigate the impactor retention ratio during the bombardment history of the Moon, a detailed quantitative study of the consequences of hypervelocity impacts of cosmic bodies of given mass, velocity, and angle of incidence are required. Modelling using shock physics codes constitutes the most accurate approach to estimate mass of projectiles retained in the crust and mantle as the result of a collision. However, given the large number of collisions on the Moon, it is impossible to model each impact event individually. A parameterization of the relationship between the properties of an impact event (projectile diameter *d*, impact velocity *v*, and impact angle *a*) and the resulting impactor retention ratio, *f*, is required. The impactor retention ratio for individual impacts with different impact angles and relatively low impact velocities ($v$ ~ 3.0-5.0 km s$^{-1}$) have been estimated from laboratory impact experiments (for example, *refs. 44-47*). Whether the results from the terrestrial laboratory experiments can be extrapolated to impactor sizes of several hundreds of kilometers in diameter and to much higher velocities in a low gravity regime is questionable. Therefore, we have carried out a systematic modeling study to determine the impactor retention ratio under realistic impact conditions on the Moon to develop parameterizations of the retention ratio as a function of impactor size, angle, and velocity on the Moon.

We use the shock physics code iSALE-3D[19] to conduct a series of three-dimensional numerical models of impacts with projectile diameters $d$ ranging from 10 to 560 km. We approximate the Moon as a 3,500-km-diameter sphere with a 700-km-diameter iron core. The Analytic Equation of State (ANEOS, *ref. 48*) for dunite[49] and iron[48] are used to describe the thermodynamic behavior of lunar mantle and core, respectively. The initial densities of dunite and iron are assumed to be ~ 3.0 g cm$^{-3}$ and ~ 7.8 g cm$^{-3}$, respectively. For undifferentiated impactors ($d$ < 300 km), we assume a dunitic composition, similar to the lunar mantle, whereas for differentiated projectiles ($d$ > 300 km), we assume the existence of an iron core accounting for ~ 30% of the total mass of the impactor[50]. The present version of iSALE-3D does not allow considering different martials that may mix upon impact. This limitation is not critical for the Moon, because the sizes of the impactors considered in this study are too small so that the projectiles cannot reach the lunar core-mantle boundary. However, for a differentiated impactor material mixing of the impactor's core and mantle within the mantle of the Moon obviously occurs. To represent this mixing despite of the code limitations, we approximate the differentiated impactor by a homogenous dunitic sphere with a radius a little larger than the real one in order to preserve the total mass. We identify with "core material" that within a central sphere with a radius similar to the metal core in the differentiated impactor and then track the fate of this material by tracers. This approximation is plausible because the approximated projectile preserves the size of the metal core and the total mass of the real differentiated impactor, and simultaneously, the size of the impactor does not change significantly. For example, for a differentiated impactor with radius of 300 km, its metal core is ~ 160 km in radius, representing ~ 30% of the total mass (e.g., ~ 4.5 x 10$^{20}$ kg). However, for an approximated dunitic sphere with the same mass, the radius is ~ 319 km. The difference on the radius between the differentiated and approximated impactor is small and decreases for smaller differentiated impactors.

In all models the impactor is resolved by 20 cells per projectile radius (CPPR), which is considered to be sufficient for accurate predictions of the impactor retention ratio[19, 51]. However, for small impactors (for example, $d < 50$ km) the high resolution causes the problem that the entire Moon cannot be modeled with the same resolution, given that the computational domain has a maximum total extent of 700 x 350 x 350 cells. Previous models indicated that the curvature of the Moon does not affect the cratering and ejection process for impactor diameters < 50 km[52]. Therefore, for small impactors, we approximate the Moon by a planar target with the same resolution for the impactor (20 CPPR). The high resolution models are very demanding on computational resources for each simulation. Each simulations lasts 1-2 months on a 96-core parallel CPU computer cluster. In total, we ran ~ 200 oblique impact models with impact velocities $v$ = 10, 15, and 20 km s$^{-1}$, and impact angles $a$ = 20º, 30º, 45º, 60º, 70º, and 80º with respect to the lunar surface. These ranges are considered to represent the most likely impact velocities and angles during the Moon's impact history[20,21,53]. The parameters used in the model are listed in Extended Data Table 1.

Our simulations aim at determining the mass fraction of impactor material deposited on and in the Moon. For each individual impact simulation, we track the impactor material by using Lagrangian tracers in iSALE, which are initially placed in the center of each computational cell and represent the mass of the material originally in the considered cell throughout the simulations (see, for example, *ref. 54*). We record the number of impactor tracers that are not ejected or are ejected from the crater with a velocity ($v_e$) smaller than the Moon's escape velocity (e.g., $v_e < 2.4$ km s$^{-1}$) and consider that the mass represented by these tracers is retained on the Moon.

Vaporization of the impactor material (i.e., dunite) requires a high impact-induced peak shock pressure (> ~ 186 GPa, *refs. 49, 55, 56*) and thus, high-velocity impacts[57].

Previous works (*refs*. 17, 18, 57) indicate that the amount of vaporized impactor is limited to a small fraction of the projectile mass for asteroidal impacts with velocities smaller than 20 km s$^{-1}$ and the vaporized volume of the impactor decreases with the impact angle relative to the surface. For the impacts on the Moon, the average velocity is relatively low, in particular for the early large impacts[58], therefore, the impactor vaporization has little influence on the impactor retention ratio for oblique impacts on the Moon[18]. In our simulations, we applied a density cut-off to erase low density (< 1 kg/m$^3$) materials that tend to expand at a relatively high velocities slowing down the computation time significantly. These low density cells usually represent vaporized impactor and target materials. The fraction of the removed material relative to the total amount of impactor material that is either retained on the Moon or escapes the lunar gravity field is usually negligible. It has been proposed recently that impactor metal cores can vaporize with a shock pressure ~ 507 GPa during high-velocity impacts, allowing iron metal to escape from the lunar surface and reducing the accreted mass of impactor material[17]. However, to be effective, this shock pressure requires extremely high impact velocities. For the impact velocities considered in this work, the amount of vaporized impactor core is very small and therefore, can be neglected.

Our simulations indicate that the retention ratio varies significantly with the impact angles (Fig. 1). High-angle impacts lead to a higher retention factor than low-angle impacts, in agreement with the results from the laboratory impact experiments (for example, *refs*. 44-47, 59). This result is not difficult to understand intuitively. For low-angle impacts, most of the impactor is sheared off at high velocity upon impact on the lunar surface, so that such impact scenarios do not add significant amount of impactor mass to the Moon (see also *refs*. 57, 59). For high-angle impacts, most of the impactor material is compacted and forms a veneer lining the crater wall of the transient crater, where it is eventually mixed into the lunar crust or mantle. In addition, for low-angle impacts ($a < 45°$), the cratering efficiency (the ratio of excavated mass to impactor mass) decreases with increasing impactor-to-target size ratio ($x$) owing to the dramatically reduced coupling between impactor and target on a curved surface (at fixed impact angle)[60]. Therefore, material from a large impactor escapes more easily than that from a small one. However, for steep impacts ($a > 45°$), the crater efficiency does not vary significantly with the impactor-to-target size ratio; as a consequence, the fraction of the impactor accreted on the Moon varies insignificantly with the size of impactors at fixed and large impact angles. Moreover, the impact velocities in our simulations are significantly larger than the Moon's escape velocity (~ 2.4 km s$^{-1}$). Most ejected impactor material has velocities much higher than the Moon's escape velocity during the course of crater excavation. Thus, impacts with different velocities (at fixed impact angle) follow qualitatively similar trends regarding the impactor retention ratio (see Fig. 1). Note that, for low-angle impacts, where a significant amount of impactor material is lost to space, N-body simulations have shown that most of the escaped material will re-hit the Earth-Moon system after a few orbits, with a higher (> 90%) probability of hitting the Earth[61], rather than the Moon, due to Earth's larger cross-section. Therefore, we consider the fraction of the escaped projectile material that subsequently collides with the Moon provides a negligible contribution to the lunar impactor retention ratio.

In order to derive a simple parameterization of the retention ratio, $f$, as a function of impactor diameter, impact angle and velocity, we fit the retention fraction from our simulations as a function of impactor-to-target size ratio ($x$). We find that the impactor retention ratio decreases while $x$ increases, following an exponential function ($f = a*\exp(-b*x)$, here, $a$ and $b$ are the fitted parameters) at fixed impact velocity and angle (see Fig. 1). The parameters of the fitted exponential function for the oblique impacts are listed in Extended Data Table 2.

The Moon experienced a long-term thermal evolution since its formation (e.g., *refs. 7, 62*). The changing thermal gradient of the Moon has a significant effect on the formation of large-scale impact basins[63,64]. However, whether the thermal gradient of the Moon affects the impactor retention ratio is unknown. To quantify this effect, two plausible thermal profiles (TP1 and TP2, see Extended Data Figure 1) are employed to represent the general thermal condition of early (warm) and late (cold) Moon. TP1 (cold) has a temperature gradient of 30 K km$^{-1}$ below the lunar surface, and decreases slowly following an adiabatic gradient (0.5 K km$^{-1}$) from temperatures in excess of 1,300 K[65]. TP2 (hot) has a temperature gradient of 50 K km$^{-1}$ in the upper 20 km, then follows the solidus of dunite within a depth range of 20-350 km, and remains constant at a temperature of 1,670 K for depths larger than 350 km (see *ref. 66*). These two temperature profiles are considered as an envelope of possible thermal conditions of the Moon during its impact history[7,62-67]. The result from our simulations indicates that the impactor retention ratio is insensitive to Moon's thermal profile (see Extended Data Figure 2 for the retention ratio of impactor with $d$ = 210 km, $v$ = 15 km s$^{-1}$ and impact angles from 15° to 80° on the Moon with the temperature profile of TP1 and TP2, respectively). The temperature, and thus, the rheology of the target plays a minor role during the excavation phase of the crater, when ejection of target and impactor takes place. Subsequently, during the crater modification phase, the thermal state and its softening effect on material strength is much more important; however, at such a late stage of crater formation no further material, in particular no impactor material is ejected anymore[68].

We also investigate where the *retained* impactor material stays on the Moon. For each individual impact, we distinguish two cases for the *retained* impactor materials: (1) ejected (with ejection velocities $v_e$ < 2.4 km s$^{-1}$, the escape velocity of the Moon) but deposited beyond the transient crater on the surface of the Moon and (2) trapped within the transient crater. According to our simulations, ~ 90% - 99% of the *retained* impactor materials remains within the transient crater and only ~ 1%-10% of the *retained* impactor materials are ejected and deposited beyond the transient crater, depending on the impact angle (see Extended Data Figure 3 for the percentage of *retained* impactor located within the transient crater). These values are broadly consistent with the results from the laboratory impact experiments, which also show that the vast majority of impactor material stay inside the transient crater (e.g., *refs. 45-47*). In addition, these values are supported by the detection of impactor relics in the Apollo samples (e.g., *refs. 69, 70*). These impactor relics were thought to originate from Imbrium or Serenatatis ejecta.

**Impactor size-frequency distribution and impact flux used in Monte Carlo modeling.**

The size-frequency distribution (SFD) of projectiles colliding with the Moon can be obtained from the impact cratering records on the lunar surface[4] as done in previous studies (for example, *refs. 3, 58*). More specifically, we convert the crater size-frequency distribution on the lunar surface[3] to the projectile size-frequency distribution by using the crater to transient crater[71,72] and the transient crater to projectile size scaling laws[22] with a prescribed distribution of impact velocities[20] and probability of impact angles[21] on the Moon. We then generate random impactors to reproduce craters and basins as we observe on the lunar surface. The number of large objects in the population can conceivably be constrained by the large basins observed on the lunar surface, provided that these impacts took place after the formation of the Moon's crust. For basins on the lunar surface, the largest confirmed impact structure is the South Pole-Aitken basin with a size of ~ 2,500 km. However, it is conceivable that the traces of some very large objects that struck the early Moon are not preserved (e.g, Procellarum with a diameter of 3,200 km, *ref. 73*) or that such events did not leave behind obvious surface or gravity expressions[74]. Our model predicts that the Moon was hit by more than one South Pole-Aitken-size (or larger) impactors[66]. Although the basin record of the Moon only shows one basin of this size[30,31,73], the prediction of our model is probably not invalidated because large-scale impact structures on the early Moon may relax quickly[32] and therefore may now be invisible on the current surface of the Moon.

Concerning the impact rate, we considered two classic scenarios. The nominal model assumes that the lunar bombardment decayed monotonically since the time of formation of the Moon[4]. The other one is the so-called sawtooth bombardment scenario with a discontinuous impact bombardment profile[3]. *Morbidelli et al.*[9] proposed a new scenario for the Moon's impact history based on numerical simulations of planetesimals left-over from the period of terrestrial planet formation. The production function obtained in this new scenario is very similar to that proposed in *Neukum et al.*[4], but it is curved upward mostly before 4.35 Gyr ago. Therefore, the result on the HSE retention age of the lunar crust and mantle would not change significantly, and consequently, we do not consider the new impact scenario in this work. The production functions considered in this work are shown in Extended Data Figure 4. For the nominal case, we extrapolate the curve up to 4.50 Gyr ago, similar to earlier works (e.g., *refs*. *3,58*). Both curves start at 4.50 Gyr ago, which is the most commonly assumed time of Moon formation[25].

**Impactor mass accreted to the Moon.**

In the Monte Carlo simulation, we varied the starting time from which we calculate the mass accreted to the Moon to the present day. With the assigned impactor flux (e.g., *ref. 4*), we estimated the total number of impactors between the starting time to the present day. For each individual impact within a time interval, the transient crater diameter and depth ($d_{tr}$) was estimated according to the scaling laws derived from laboratory impact experiments[75] and numerical modeling[52]. If the depth of transient crater ($d_{tr}$) is larger than the average crustal thickness, the impact is assumed to penetrate through the crust and to transport the *retained* impactor mass trapped within the transient cavity directly into the lunar mantle; instead, the fraction of *retained* impactor that was deposited beyond the rim of the transient crater is assumed to mix with the crust. In all other cases, the total *retained* impactor mass is mixed with the crust. The latter assumption is reasonable because small projectiles do not penetrate the crust and therefore cannot transport the *retained* material to the mantle. For large objects that punch through the crust, the impact produces large volumes of impact melt[76,77] forming a melt sheet that connects to the mantle[78]. The fraction of *retained* impactor materials trapped within the transient cavity can be directly transported into the mantle through the impact melt pool[78,79], leaving just the fraction of *retained* material deposited beyond the transient cavity to mix with the crust. According to our modeling (see Extended Data Figure 3), we assume in our Monte Carlo calculations, for simplicity, that ~ 96% of the *retained* impactor material is trapped within the transient crater and transported into the mantle, and the remaining ~ 4% of the *retained* impactor material is deposited beyond the transient crater rim and mixed with the crust. This assumption is reasonable because our results are not sensitive to the fraction of the *retained* impactor material mixed with the crust (see the following discussion).

We use the average crustal thickness of 34 km and 43 km derived from the GRAIL data[23] as a lower and upper limit for the crustal thickness of the Moon. As the crust is thought to form very soon after the Moon formation[24,27], we assume that the crustal thickness is constant over time in our calculation. For each starting time, we repeat the Monte Carlo procedure millions of times to address the stochastic variability intrinsic to the bombardment process and record the masses accreted to the crust and mantle, respectively. Note, for differentiated impactors ($d > 300$ km, *ref. 50*) for which the impactor's core is accreted to the Moon (see Extended Data Figure 5a), we assume that the *retained* material of the impactors and their (*retained*) HSEs are fully merged into the lunar crust and/or mantle. Although these assumptions are probably true for small impactors, larger objects (e.g., $d \gg 300$ km), to some extent, are thought to merge a fraction of their cores with the core of the Moon directly during the impact. We analyzed our impact simulations $d$ = 300-560 km and found that none of the accreted impactor's cores merge with the core of the Moon directly. However, as the impactor core is approximately assumed as the center region of a homogenous impactor with the same material and density as the lunar mantle, rather than the denser metal material that can reach the Moon's core via impact, the fraction of the *retained* core merged with the Moon's core is underestimated to some extent in our simulations. Several additional impact models with $d$ = 640-960 km show that only impacts with high impact angles (> $70^\circ$) have a small fraction of the *retained* core material directly merged with the Moon's core. This fraction increases with the impactor diameter and impact angle, but never exceeds ~10%. However, in our Monte Carlo simulations, such large impactors are extremely rare in the Moon's impact history and always have low impact angles (e.g., < $30^\circ$), therefore, the fraction of accreted impactor core material directly merged with the Moon's core is considered to be insignificant. In our calculation, we record the total mass of the bodies accreted to the Moon. However, for low-angle impacts, for which the impactor's core is not accreted (see Extended Data Figure 5b), we do not record any mass (from the mantle and crust of the impactor) accreted to the Moon because the HSE concentration in the crust and mantle of differentiated impactors are assumed to be very low. Fig. 3 shows the accreted mass into the crust and mantle for the crustal thickness of 34 km and 43 km for the impact flux from *Neukum et al.*[4]. Our results indicate that the thick lunar crust increases the proportion of accreted materials into the crust. However, the accreted masses into the crust are within the mass ranges inferred from the crustal HSE budget. We also did a similar calculation with the assumption that ~ 10% of the *retained* impactor is deposited beyond the transient crater and is mixed with the crust (see Extended Data Figure 6). Our results show that the accreted mass into the crust and mantle are still within the masses inferred from the HSE budgets. Impactors accreted to the crust and mantle reach the masses inferred from the HSE concentrations if HSE retention began at ~ 4.35 Gyr.

HSEs were known to be removed from the mantle by metal segregation during core formation. Because the Moon's core should have formed quickly after lunar formation[25], the HSEs in the lunar mantle and crust have been used to constrain the amount of chondritic material hitting the Moon since its formation. With an average impactor retention ratio of 0.6, 2-3 times larger than the value derived in this work, the total mass accreted by the Moon exceeded that predicted by the HSEs by almost an order of magnitude, if the impact flux of *Neukum et al.*[4] is adopted (see *ref. 3*). As a consequence, an impact flux was designed with a sawtooth-like profile (see Extended Data Figure 4) to reconcile the formation of numerous craters and basins at the time of the late heavy bombardment (i.e., younger than 4.1 Gyr) with the total mass accreted to the Moon inferred from the HSEs (see details in *ref. 3*). Although our new results indicate a lower impactor retention ratio, ~ 2-3 times lower than the previous considerations[1,3], it does not reconcile the impact flux of *Neukum et al.*[4] with the lunar HSE budgets if the lunar crust and mantle retained HSEs since the time of lunar formation (e.g., ~ 4.50 Gyr ago). Only if the mantle started to retain HSEs since 4.35 Gyr (i.e., ~ 150 Ma after the Moon formation), as proposed in (*9*) and in this work, the lunar HSE budget can be explained with the impact flux of *Neukum et al.*[4]. In this case the sawtooth profile of the production function is no longer needed. To the sequestration of the HSEs accreted before 4.35 Gyr into the lunar core likely involved sulfide segregation to the lunar core during the end of LMO crystallization and mantle overturn (e.g., *ref. 9*).

**Average impactor retention ratio for the Earth.**

The Earth's escape velocity is almost 5 times higher than that of the Moon (11.2 km versus 2.4 km/s). The total mass of escaping impactor material is ~ 5-25% for impact angles of 30º and 15º, correspondingly, and is negligible at higher impact angles[80]. Thus, the impactor retention ratio for the Earth is only slightly less than 1.

**Data availability**

The data that support the findings of this study are available from the corresponding author on request.

**Code availability**

At present, the iSALE is not fully open source. It is distributed on a case-by-case basis to academic users in the impact community, strictly for non-commercial use. Scientists interested in using or developing the iSALE should see http://www.isale-code.de for a description of application requirements. The Monte Carlo code used in this work is available from the corresponding author on request.

**Extended Data Figure 1 | Thermal profiles of the Moon.** Two possible thermal profiles of the Moon are used to test the effect of the temperature profiles on the impactor retention ratio.

**Extended Data Figure 2 | Effect of thermal profiles of the Moon on the impact retention ratios.** Note, the impactor retention ratios were calculated for oblique impacts with $d$ = 210 km and $v$ = 15 km s$^{-1}$. The impact angles were varied from 15$^o$ to 80$^o$. For TP1 (orange points), impactor retention ratios for impacts with angle of 40$^o$, 50$^o$, and 90$^o$ are also plotted. Note, TP1 and TP2 in the figure represent the temperature profiles used in this work as shown in Extended Data Figure 1.

**Extended Data Figure 3 | Fraction of *retained* impactor material deposited within the transient crater for all simulations.** In our simulations, this fraction is within 0.9 – 1.0 for impact angles larger than 20 degrees (relative to the lunar surface). For large impacts ($d$ > 100 km) with impact angles smaller than 20 degrees, the fraction of *retained* material within the transient cavity is lower than 0.9. The dashed line represents the fraction of 0.96 (see Fig. 3) we use in our calculations for simplicity. The dotted line represents the fractions of 0.90 used in the calculations as shown in Extended Data Figure 6. The numbers in the legend represent the impactor diameter (D) and impact velocity (V), respectively.

**Extended Data Figure 4 | Lunar impact fluxes.** The differential number of lunar crater > 20 km (N$_{20}$) as a function of time and per unit surface for the production functions discussed in the text.

**Extended Data Figure 5 | Two scenarios for a differentiated impactor hitting the Moon.** When the differentiated impactor core is accreted to the Moon (panel **a**), we record the total impactor mass accreted to the Moon. However, for the scenario where the

impactor's core was not accreted to the Moon (panel **b**), we do not record any mass accreted to the Moon. The reason is that the simplified impactor is assumed to have uniformly distributed HSEs, while in reality the HSEs of a differentiated impactor should almost entirely be into its core. The arrow represents the impact direction and the lines show the extent of interaction of the impactor core with the Moon.

**Extended Data Figure 6 | The cumulative impactor mass that hit (blue) and accreted (purple) on the Moon.** Similar to Fig. 3, but we assume that ~ 10% of the retained impactor material are deposited beyond the transient crater and mixed with the crust. The crustal thicknesses are 34 km (**a**) and 43 km (**b**), respectively.

**Extended Data Table 1 | Model parameters used for iSALE-3D simulations.**

[a] See *ref. 49*.
[b] See *ref. 48*.
[c] See *ref. 54*.
[d] See *refs. 52, 64, 66 and references therein* for a description of the strength model parameters and their implementation in iSALE.
[e] See *ref. 66 and references therein* for a description of the strength model parameters and their implementation in iSALE.

**Extended Data Table 2 | Parameters of the exponent functions ($f = a*\exp(-b*x)$) for the fit of impactor retention ratio derived from the oblique impact simulations (see Fig. 1)**. Here $x$ represents the ratio of the diameter between the diameter and the Moon.

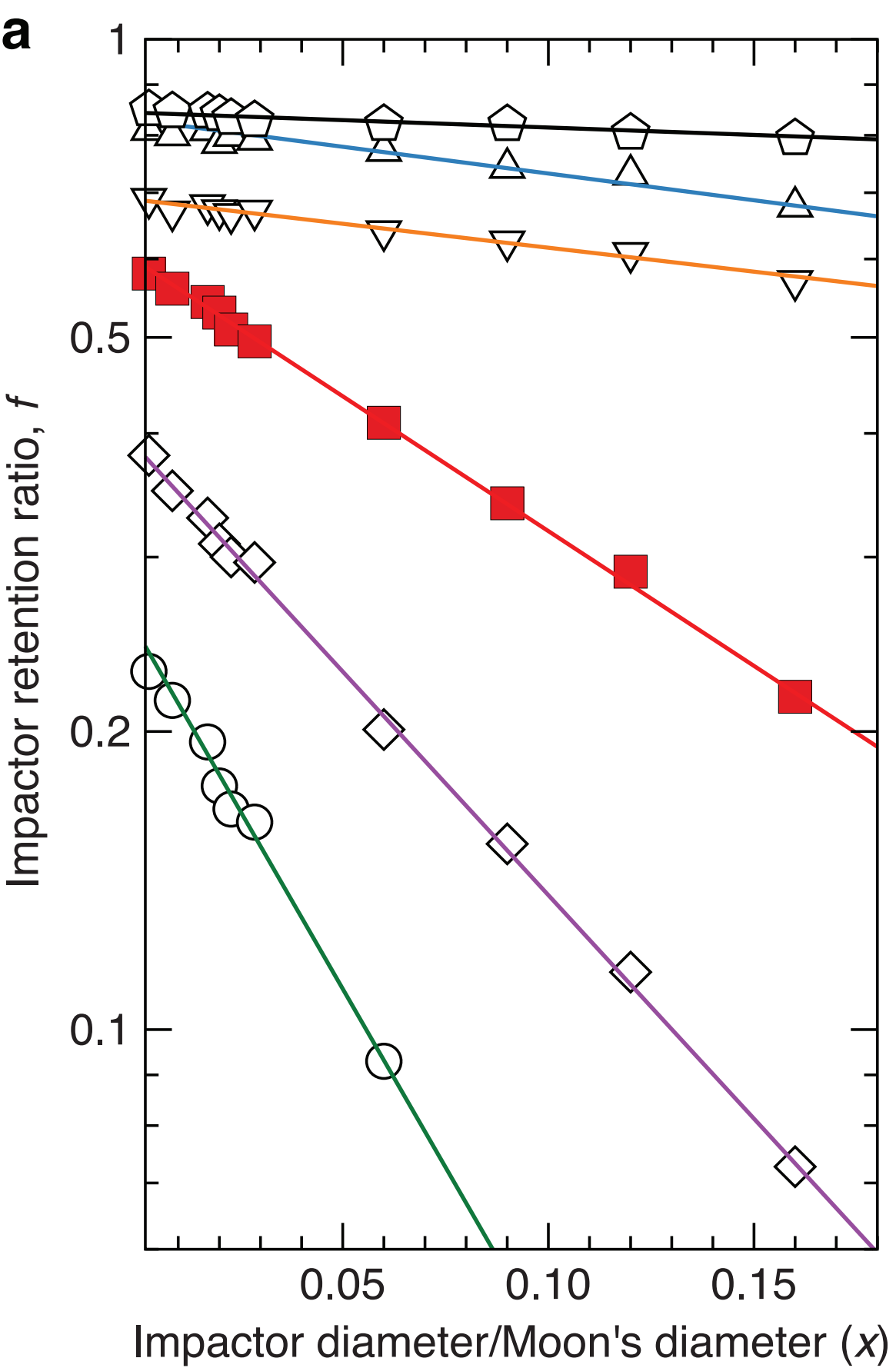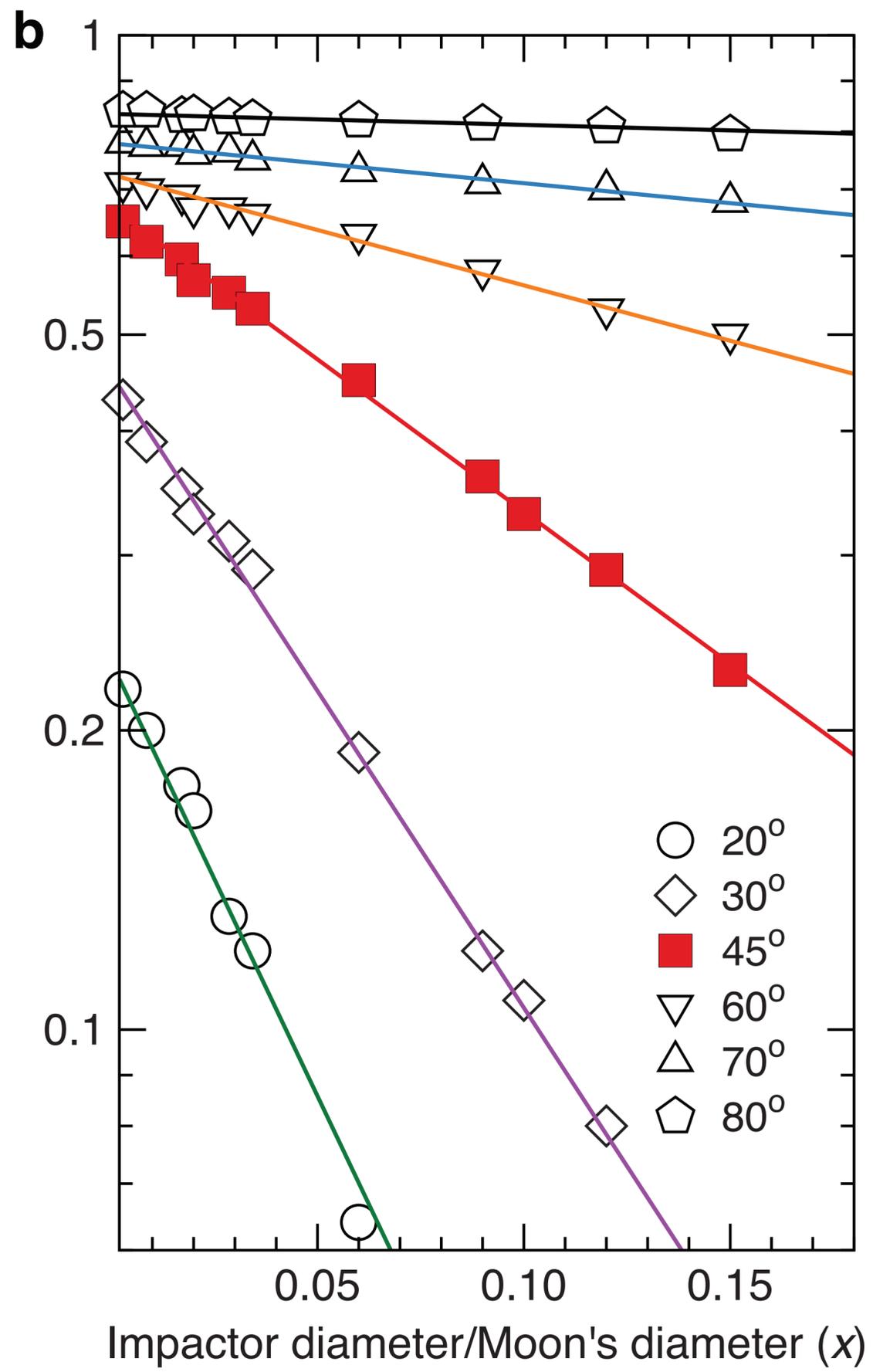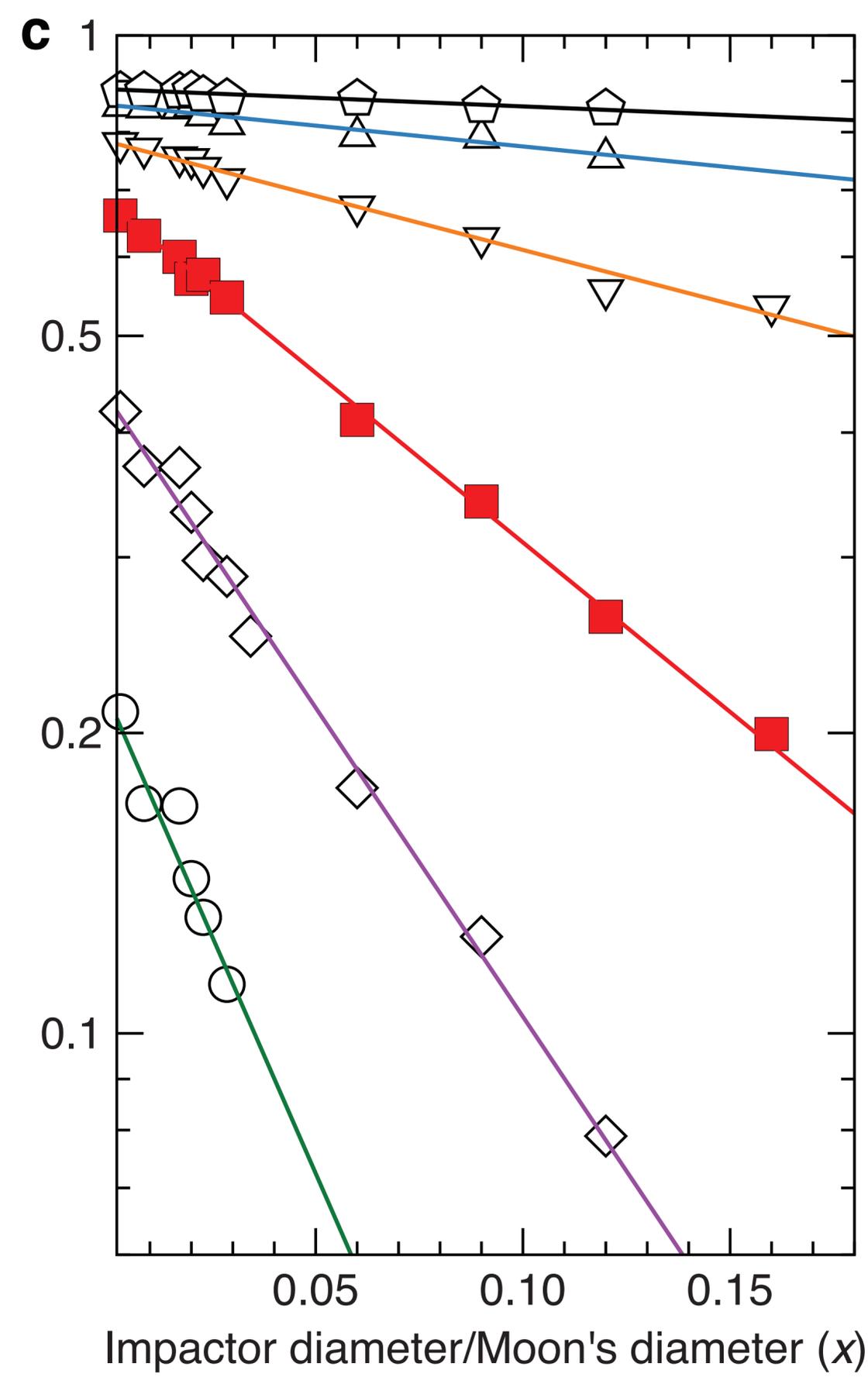

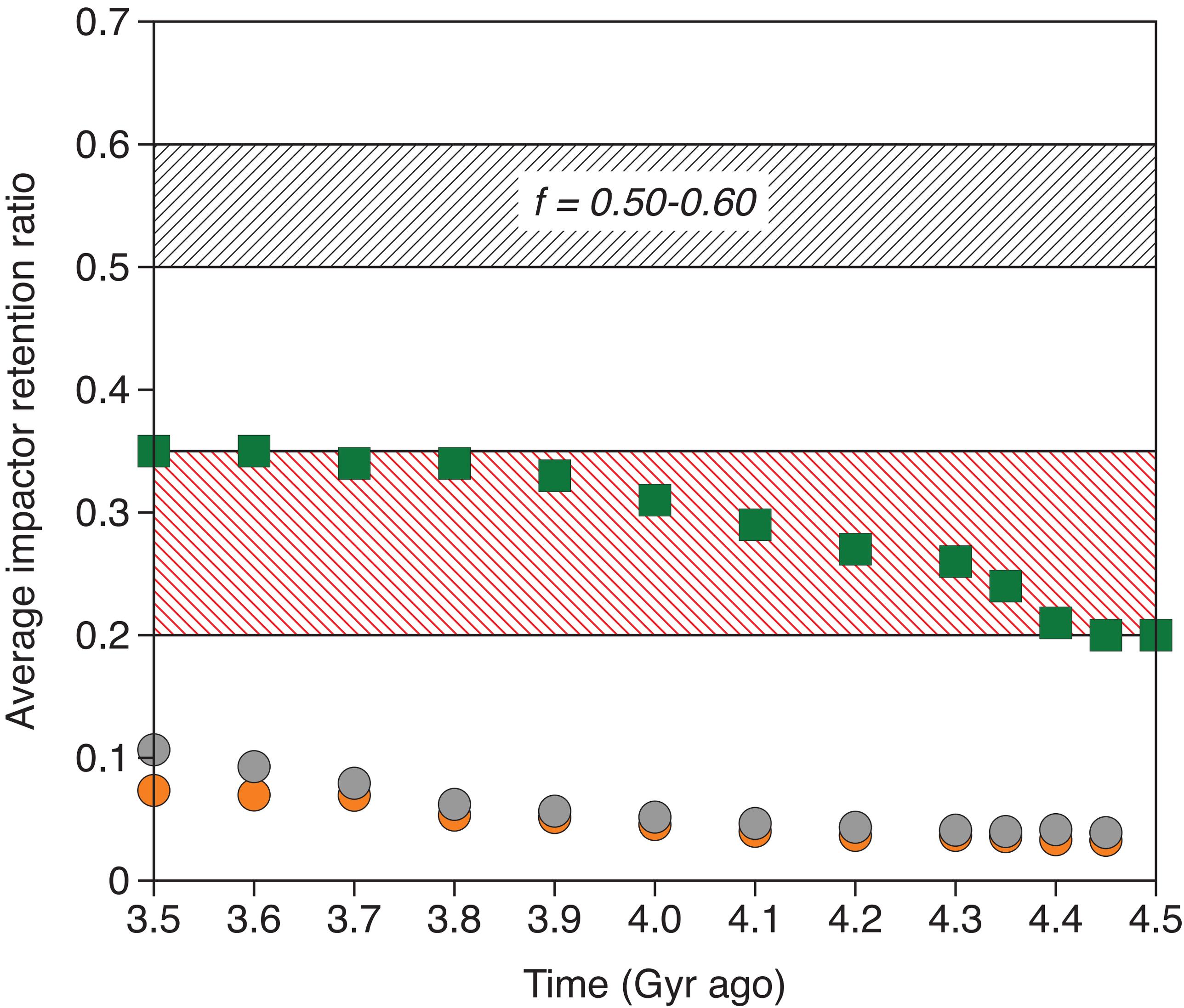

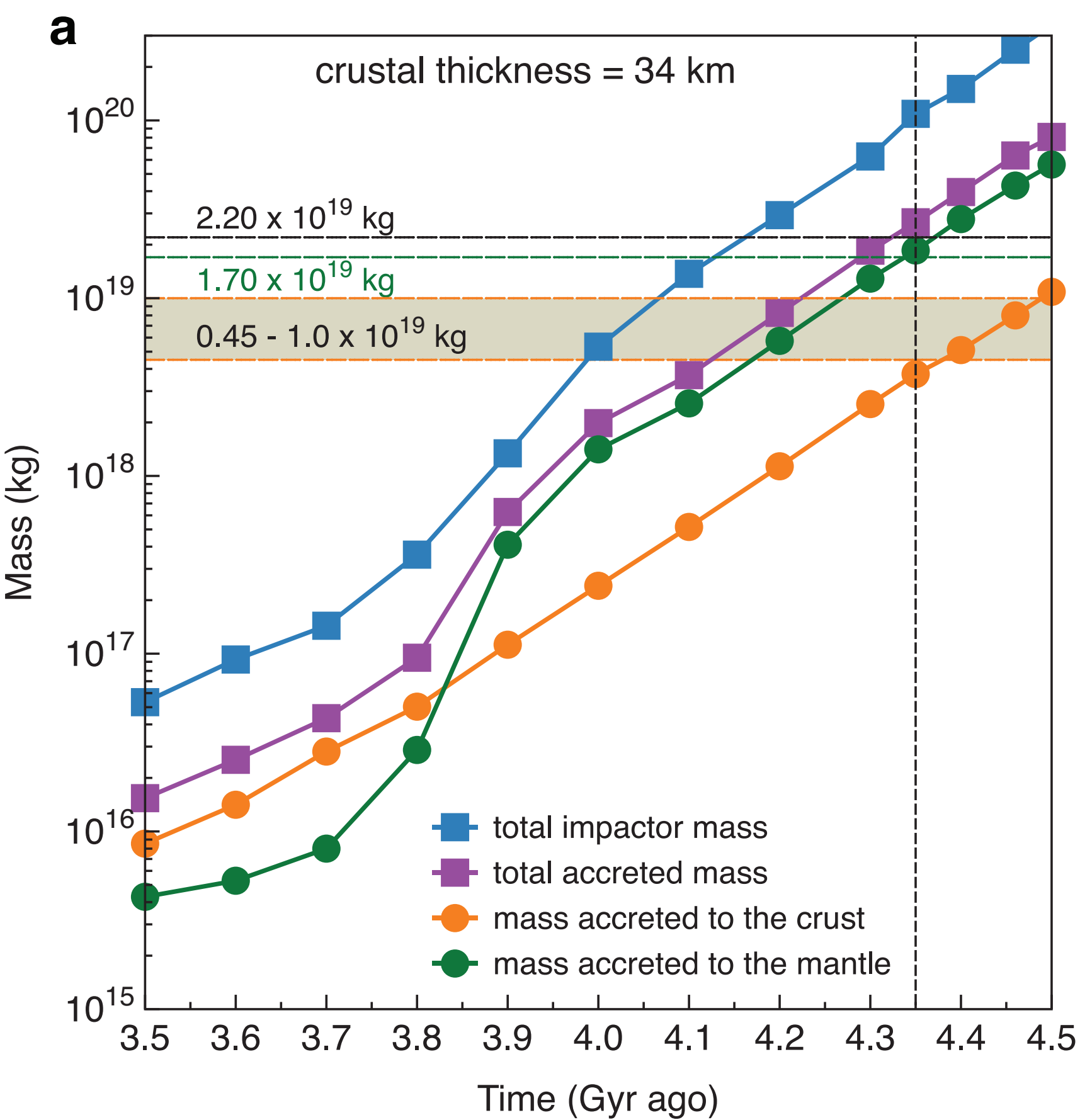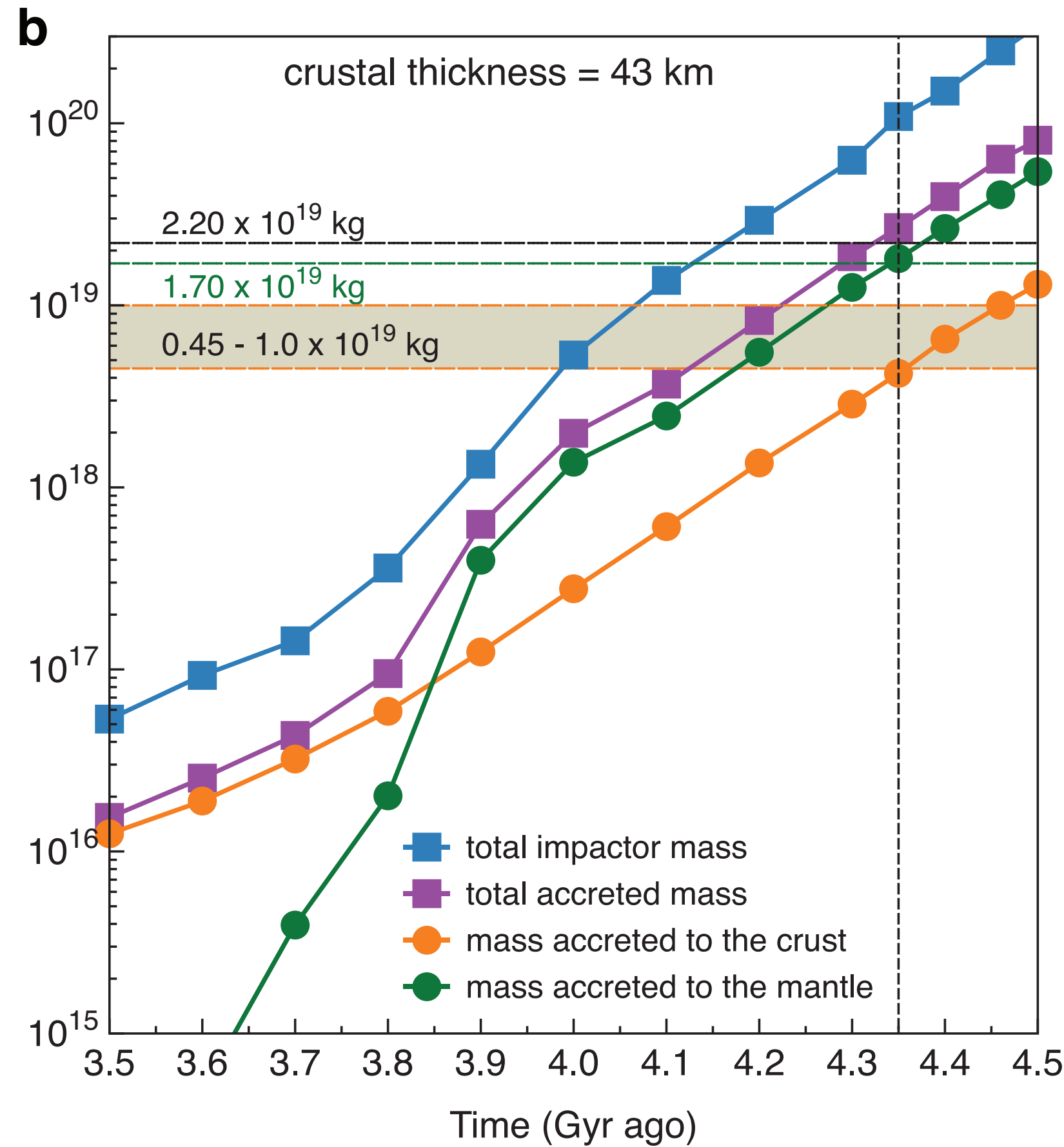